\documentclass[sn-mathphys]{sn-jnl}% Math and Physical Sciences Reference Style
{}

\jyear{2022}%

\begin{document}

\title[Modelling photo-evaporation in planet forming discs]{Modelling photo-evaporation in planet forming discs}

\author*{\fnm{Barbara} \sur{Ercolano}}\email{ercolano@usm.lmu.de}

\author{\fnm{Giovanni} \sur{Picogna}}\email{picogna@usm.lmu.de}
\equalcont{These authors contributed equally to this work.}

\affil*{\orgdiv{Universitäts-Sternwarte}, \orgname{Ludwig-Maximilians-Universität München}, \orgaddress{\street{Scheinerstr. 1}, \city{München}, \postcode{81679}, \state{Bayern}, \country{Germany}}}

\abstract{Planets are born from the gas and dust discs surrounding young stars. Energetic radiation from the central star can drive thermal outflows from the discs atmospheres, strongly affecting the evolution of the discs and the nascent planetary system. In this context several numerical models of varying complexity have been developed to study the process of disc photoevaporation from their central stars. We describe the numerical techniques, the results and the predictivity of current models and identify observational tests to constrain them.}


\maketitle

\section{Introduction}\label{sec1}

The formation and evolution of planetary systems is strongly coupled to the evolution and final dispersal of the protoplanetary discs in which they form. This is driven to a large extent by irradiation from the central star, which provides heating, ionization (thus magnetic field coupling) and can trigger thermal outflows that disperse the disc material. 

The dispersal of the gas disc imposes a final timescale within which giant planets must form and is of crucial importance for the final architecture of planetary systems. Photoevaporation operates by first opening a gap in the disc, which allows the inner disc to drain onto the central star while the outer disc is dispersed from the inside out. This provides a natural parking mechanism for migrating planets \citep{Alexander2012,ErcolanoRosotti2015,Jennings2018,Monsch2019} as well as influencing the post-disc dynamical evolution of the orbits \citep{Moeckel2012,Liu2022}. The final distribution of the semi-major axis of giant planets in a population is extremely sensitive to the mass-loss profile of the disc \citep{ErcolanoRosotti2015, Jennings2018}.

The role of photoevaporation in the early phases of planet formation, e.g. the formation of planetesimals by the streaming instability is instead debated \citep{Ercolano2017,Carrera2017}. For the streaming instability to occur a high solid to gas ratio is needed. Photoevaporation is usually invoked as a mechanism that preferentially removes gas, thus helping to locally increase the dust-to-gas ratio. While \citep{Ercolano2017} test this hypothesis and find that the effect of photoevaporation on the total planetesimal mass is negligible, the opposite conclusion is obtained by \citep{Carrera2017}, using a similar method. The reason for this discrepancy lies in the different photoevaporation mass-loss profiles assumed in these two different works.  

Realistic mass-loss profiles are also a key ingredient in planet and disc population synthesis models \citep[e.g.][]{Emsenhuber2021a,Emsenhuber2021b,Manara2018}, while a reliable disc surface density evolution consistent with the thermodynamical wind structure is crucial for the interpretation of observations of transition discs \citep{Rosotti2015,Owen2016,Ercolano2017b,Schaefer2022} and gas emission lines and dust observation \citep{ErcolanoOwen2010,ErcolanoOwen2016,Weber2020,Franz2022a,Franz2022b}.

In this context several theoretical and numerical models of the photoevaporation process in planet-forming discs have been developed over the years, having as the main objective the determination of accurate spatially-resolved mass-loss rates.  

Recent theoretical and observational advancements \citep[see reviews][]{ErcolanoPascucci2017,Pascucci2022} which point to disc winds of magnetic nature (MHD winds) in addition to thermal (photoevaporative) winds have added to the urgency of developing quantitatively predictive thermal wind models in order to distinguish them in the observations \citep{Weber2020,Ricci2021}. This is of paramount importance to assess the relevance of MHD winds to angular momentum extraction from discs.

In this paper, we will review the current stand of numerical photoevaporation models. Special attention will be given to the applicability of the different models, as well as to the divergence of the results. Present and future observational tests of the models are also reviewed. 

\section{Numerical modelling}

Disc photoevaporation is a coupled problem in radiative transfer, thermochemistry, hydrodynamics and dust dynamics. Until recently, it was deemed impossible to combine all these effects into comprehensive numerical simulations, but in recent years it has been shown that the codes and numerical capabilities are available \citep{WangGoodman2017,Nakatani2018a}, though still with significant limitations in the radiative transfer.
Theoretical photoevaporation models have been developed for one or a combination of three different portion of the stellar spectra that are capable of driving a disc thermal wind: the far-UV (FUV) $6 \mathrm{eV} < h\nu < 13.6 \mathrm{eV}$; extreme-UV (EUV), $13.6 \mathrm{eV} < h\nu < 100 \mathrm{eV}$; and X-rays, $100 \mathrm{eV} < h\nu < 10 \mathrm{keV}$ (in particular its soft component $100 \mathrm{eV} < h\nu < 1 \mathrm{keV}$). Each of them presents its characteristic features and numerical challenges, so it is important to distinguish between them.

\paragraph{EUV radiation}
EUV photons are defined as energetic enough to ionize atomic hydrogen from its atomic state ($h\nu > 13.6 \, \mathrm{eV}$).
Since the ionization cross-section for neutral hydrogen is $10^{-17} \mathrm{cm}^2 \mathrm{atom}^{-1}$, an EUV photon will be absorbed after passing a neutral column density of $N_H = 10^{17} \, \mathrm{cm}^{-2}$. %This translates generally to a penetration depth of $\sim 10^{20}\, \mathrm{cm}^{-2}$ \citep{Ercolano2009} since some of the material at equilibrium will be already ionized close to the star. -- i would avoid saying this as it depends on the luminosity
The interaction between a EUV photon and a hydrogen atom is straightforward. The photon is absorbed, producing an ionized hydrogen atom and thermal energy. Typical temperatures in ionized regions are $\sim 10^4 \, \mathrm{K}$. The production of the electron-ion pair is offset by recombinations. Those to the electronic ground state produce a photon with energy larger than 13.6 eV which can further ionize the medium, producing a diffuse field. The recombinations to an excited state instead lead to the destruction of the ionizing photon.
In the case of irradiation by a central star of a primordial disc with typical scale heights, photons from the central star generate a hot hydrostatic atmosphere of $\sim 10^4 \, \mathrm{K}$ above the inner disc. It is then the diffuse field of recombination photons from this atmosphere that irradiates and ionizes the outer disc. This problem has to be solved as a 2D radiative transfer, which takes into account the direct and diffuse ionizing fields while solving the photoionization problem.

\paragraph{FUV radiation}
FUV photons have energies in the range of $6 - 13.6$ eV, they cannot ionize hydrogen, but they effectively photodissociate molecular hydrogen, CO, and other molecules, and ionize Carbon. 
%FUV photons are produced in active chromospheres and they may penetrate accretion columns, and HII regions created by the EUV flux, reaching typical column densities up to $N_H\sim10^{21}-10^{23}$ cm$^{-2}$ depending primarily on dust properties. They are measured in nearby, young, solar mass stars with little accretion and typically (with great scatter) have luminosity ratios $L_{FUV}/L_{bol} \sim 10^{-3}$ or $\Phi_{FUV} \sim 10^{42}$ photons/s.
%The region affected by FUV is called the photo-dissociation region (PDR), and it is bounded by an ionization front on the one side (caused by the EUV irradiation) and a molecular layer on the other side. FUV photons pump molecular hydrogen to an excited electronic state, which is followed by fluorescence. In $85-90\%$ of cases, the result is molecular hydrogen in an excited vibrational state, while in the remaining cases, the system reaches the vibrational continuum and a hydrogen molecule is converted into two hydrogen atoms. PDR chemistry is dominated by photodissociation and photoionization by FUV photons and photoelectric grain heating. -- this is a little too general.. most people know this?
Photoelectric heating from dust grains and polycyclic aromatic hydrocarbons (PAHs) is the dominant gas heating mechanism for FUV photons. % since roughly $95\%$ of the incident FUV flux is absorbed by dust grains, and the rest by PAHs, which is then re-radiated as (IR) continuum. 
Heating is proportional to the grain surface, thus in the presence of abundant small grains or PAHs in the disc atmosphere, FUVs can heat the gas and potentially contribute to driving a thermal wind \citep{Gorti2009}.
The gas radiates only less than $1\%$ of the energy processed by the dust, and it is heated mainly collisionally by photoelectrons from small dust grains and PAHs, or by vibrationally excited H2. Thus its temperature is impacted by both dust evolution and chemistry \citep{Gorti2015,WangGoodman2017}.
FUV photons can heat a primarily neutral layer of H and H$_2$ to temperatures of order $10-5000$ K, depending on the magnitude of the flux, the density of the gas, and the chemistry. In order to correctly model this problem one has to combine a 2D radiative transfer, with a sufficiently large chemical network to account for the strong variation in density and temperature, and a proper dust evolution model.
Another complication arises from the fact that the component of the stellar FUV field due to accretion onto the central star is can dominate over the chromospheric component. This means that the FUV flux reaching the disc, and thus the wind mass-loss rates, strongly depend on the accretion rate which decreases with time and can stop completely once an inner cavity is formed. 

\paragraph{X-ray radiation}

X-rays can be divided into two main components: the soft X-rays ($100\, \mathrm{eV} < h\nu < 1 \,\mathrm{keV}$), which are absorbed at a column density of $\sim 10^{22}$ pp/cm$^2$, heating up the disc’s upper layers where they drive a thermal wind; the hard X-rays ($1\, \mathrm{keV} < h\nu < 10\, \mathrm{keV}$), which reach much deeper into the disc where dust and gas are thermally coupled, but they can increase the level of ionization (together with cosmic rays) close to the disc mid-plane thus affecting the coupling of the gas with the disc magnetic field. 
The main interaction of X-ray photons is via photoionization of (the inner shells of) atoms and molecules, which can lead to secondary ionisations (including of Hydrogen), followed by the thermalisation of the kinetic energies of primary and secondary electrons. X-ray heating produces a range of temperatures going from a few hundred K in the dense X-ray PDRs up to $10^4$ K in the disc upper layers.
%The origin of X-ray radiation from low mass pre-main sequence stars is still debated. Qualitatively, the interior dynamo of these fully convective stars erupts onto the surface where magnetic reconnection occurs. The hot X-ray emitting plasma is then confined in surface fields, with the hotter component coming from higher altitudes (as also shown in MHD simulations \citep{Cohen2017}). The 
X-ray luminosities do not depend on stellar accretion rates and in fact, they remain nearly constant with time during the first Myrs \citep{Getman2022}, 
yielding a constant wind mass-loss rate regardless of changes in the surface density of the disc. 
%suggesting that they are uncorrelated to the stellar (decreasing) radii and rotation rate, in a saturated state. This observational evidence is consistent with the scenario of an X-ray radiation coming mainly from the hot plasma at high altitudes above the stellar surface.

\subsection{Methods}

Since the realisation of the importance of disc photoevaporation, more than 30 years ago \citep{BallyScoville1982}, the problem has been tackled placing more emphasis either on radiative transfer modelling, or on the full time-dependent radiation-hydrodynamics. \\%While the second approach is very powerful, it usually comes at a cost, requiring a simplified radiative transfer to keep the problem tractable. This tension remains up to date to determine, for a given region of the parameter space, which portion of the stellar spectra is the bigger contributor to the disc wind, and thus what is the magnitude of the wind mass-loss rate.\\

The first studies of disc photoevaporation focussed on EUV photoevaporation of discs by OB stars, which are strong EUV emitters. %since these sources emit strongly in the EUV, and the nature of the interaction between EUV photons and the circumstellar material is easier to model with respect to FUV and X-ray that penetrate larger column densities.
Two lines of models started contemporarily.
The first adopting a simplified radiative transfer calculation in the Eddington approximation (reducing the three-dimensional radiative transfer problem to a “three stream” approximation coming from the star, vertically upwards, and downwards at each point in the disc) applied to a vertically hydrostatic disc \citep{Hollenbach1993,Hollenbach1994}. The second one, more focused on the dynamics \citep{YorkeWelz1993, YorkeKaisig1995}, started by developing a 2D hydrodynamic photoionization code, later improved including the effects of UV dust scattering \citep{RichlingYorke1997}. 
Both lines of studies found that the most important contribution to the wind comes from the diffuse field from a puffed-up bound inner disc and that the thermal wind is launched from the gravitational radius $R_g$\footnote{the gravitational radius is defined as the location where a gas parcel becomes unbound from the central star \citep{Hollenbach1994,Liffman2003} $R_g=GM_\star/c_s^2$. For gas at roughly 10000K in the atmosphere of a disc around a solar mass star this is roughly 5~au. The wind however is launched at the so-called critical radius $R_c \sim 0.2 R_g \sim 1~au$ \citep[e.g.][]{Dullemond2007}.  } outwards.
%heated by the EUV photons to $\sim 10^4$ K, and obtained an approximate analytical solution to their model. The agreement was also mainly driven by the nature of EUV photoevaporation, which creates an isothermal gas corona, since the EUV photons are absorbed for low column densities.
By adding the analytical prescriptions for mass-loss rates as a function of disc radius due to EUV photoevaporation as a sink term in the 1D disc viscous evolution equation \citep{Clarke2001,Matsuyama2003}, the so-called 'EUV-switch' was demonstrated (see Section~\ref{sec:results}). 

The first hydrodynamical models of EUV disc photoevaporation for T Tauri stars \citep{Font2004} were performed assuming an isothermal equation of state, and imposing the density at the base of the flow from previous results \cite{Hollenbach1994}. They found that the photoevaporative flow was launched subsonically ($\sim 0.3\, c_s$) from a much smaller radius than the gravitational radius ($R_c \simeq R_g/5$) \citep[see also][]{Dullemond2007}, which was not expected from the analytical prescription.

These studies were then followed by the first hydrostatic models of X-ray irradiation from T Tauri stars, first with a simple heating model \citep{Alexander2004}, and later with a more realistic two-dimensional Monte Carlo photoionization and dust radiative transfer code \citep{Ercolano2008b,Ercolano2009}. The latter calculations used a more realistic input spectrum which included both the EUV and X-ray components, showing that the latter completely dominates, with mass-loss rates some 2 orders of magnitude higher than previous pure EUV models. 

%finding mass-loss rates in the range $10^{-10}-10^{-8} \, M_\odot/yr$ ($\sim 20$ times larger than the predicted EUV photoevaporative rates).

At the same time, a 1+1D hydrostatic equilibrium model was also developed including EUV, X-rays and FUVs, and a comprehensive chemical network \citep{GortiHollenbach2009,Gorti2009}. These models showed that FUV radiation might be efficient at removing material from the outer disc, but only if PAHs are abundant in discs. The input spectrum used by these models was however very idealised, with a hard X-ray input spectrum, which was inefficient at heating the gas and driving the wind.

Without proper hydrodynamical calculations, all these models estimated local mass-loss rates by assuming that the local mass loss rate is roughly equal to the product of the density at the sound speed at the base of the wind, the so-called $\rho \dot c_s$ method. Without hydrodynamics, identifying the base of the wind is rather arbitrary and different authors used different approaches, which further exacerbated tension between the models.

%halting the gas resupply to the inner disc and reducing the disc lifetime. They also put to question the relative importance of X-ray/FUV stellar irradiation, as it depends strongly on the irradiating stellar spectra and the assumptions made on the dust distribution (and the importance of PAHs), leading to strong differences in the outcome of the different models \citep{Ercolano2009,GortiHollenbach2009}.

%It was than realized that an hydrodynamical model was necessary to properly constraint the photoevaporation rate, since X-ray and FUV photons penetrate much deeper into the bulk of the disc, and the exact location of this dense heated layer where the base of the wind sets, determines the wind mass-loss rates (i.e. the mass-loss rate is roughly proportional to the product of local density and sound speed at the wind base).

The first (radiation-)hydrodynamical calculation including both X-rays and EUV radiation \citep{Owen2010,Owen2012} was performed using a parameterisation of gas temperature as a function of the ionisation parameter obtained from detailed thermal and ionisation calculations \citep{Ercolano2009} performed with the {\sc mocassin} code \citep{Ercolano2003, Ercolano2005, Ercolano2008a}. The ionisation parameter is defined as $\xi = \frac{L_X}{n r^2}$, where $L_X$ is the X-ray luminosity, $r$ and $n$ are the local disc radius and the volume density of the gas. This method is generally known as the $\xi-T_e$ approach. 
%through the ionization parameter \citep{Tarter1969}.
%\begin{equation}
%    \xi = \frac{L_X}{n r^2}\,,
%\end{equation}
%which represents the ionizing energy available to a gas particle at a distance $r$ from the central star, and it determines the ionization state and thus the equilibrium gas temperature \citep{IgeaGlassgold1999}.
%Th found a mass-loss rate from a region extending out to 100 au and linearly increasing with the stellar X-ray luminosity.
%Moreover, they stressed that more importantly than the penetration depth of the different component of the spectra at a certain distance from the central star, what matters for the magnitude of the mass-loss rate is the major component heating up the disc when the gas flow is accelerated to the local gas sound speed, at the wind base.
These studies were later extended adopting the modern hydrodynamical code \textsc{pluto} \citep{Mignone2007}, and better accounting for attenuation effects in the disc using a column density dependant $\xi-T_e$ parameterisation \citep{Picogna2019,Woelfer2019,Picogna2021,Ercolano2021}.
With this improved prescription it was shown that the X-ray irradiation could reach several hundred au, and that the mass-loss rates for high X-ray luminosities reaches a plateau as more energetic photons cannot heat up lower and denser cold regions of the disc.

The advantage of the $\xi-T_e$ method is that, while being based on detailed multi-frequency thermal calculations, introduces very little computational overheads on the hydrodynamics. A drawback of this method is that the thermal calculations are performed in radiative equilibrium, and thus the contribution of adiabatic cooling cannot be accounted for. All models performed with the $\xi-T_e$ approach should therefore perform a-posteriori check to ensure that  radiative equilibrium is justified throughout the simulation domain. This has indeed been demonstrated for all the models obtained with this approach so far \citep{Owen2010, Picogna2019,Woelfer2019,Picogna2021,Ercolano2021}.

Recently some numerical experiments have been carried out to perform (a very streamlined) radiative transfer and thermochemical calculation on the fly in hydrodynamical models \citep{WangGoodman2017,Nakatani2018a}. Due to the high computational costs of this approach, only a handful of models exist with limited spectral and spatial resolution. The results obtained by these models diverge strongly from those obtained with the $\xi-T_e$ method, as discussed in Section~\ref{sec:divergence}.

%Recently, it has been shown that the capabilities of combining hydrodynamical simulations of disc photoevaporation by X-ray and EFUV radiation with consistent thermochemistry were available \citep{WangGoodman2017,Nakatani2018}, overcoming the limitations of having either a temperature prescription for the wind region or having a static disc model.
%The problem is that, even adopting similar numerical methods, the result depends strongly on the choices made for the irradiating spectra, the disc chemical composition and the dust component. It is though fundamental to understand the origin of these differences in order to understand what is driving the observed mass-loss rate.

%\begin{itemize}
%\item semianalytical approach Alexander, Clarke
%\end{itemize}

\section{Results}
\label{sec:results}

The main results from all photoevaporation models developed to date can be roughly summarised as follows: radiation from the central star heats the upper layers of protoplanetary discs, which become unbound and are centrifugally accelerated in a thermal outflow. Wind mass-loss rates peak around a specific radius of the disc known as the gravitational radius, with profiles that can be more or less extended to larger radii. No mass-loss is expected to occur from regions that are close to the star, and thus more gravitationally bound, which would require the gas to achieve very high temperatures to escape. The combination of steady mass-loss rates due to photoevaporation and a steadily decreasing accretion rate with time in a viscously accreting disc produces the so-called photoevaporation switch: Discs evolve viscously for a few million years, or until the viscous accretion rate has reached values comparable to the wind mass-loss rate, at which point photoevaporation takes over and the disc is quickly dispersed from the inside out, via the formation of a gap first and then a clear-out cavity. 

Modern hydrodynamical models of photoevaporation are performed (at least) in 2D \citep[but see][for 3D simulations including planets]{Weber2022} and thus allow only to model the disc for a limited amount of orbits, over which a steady-state solution for the outflow is (ideally) achieved. Modelling the evolution of the surface density of the disc as the gap opens and the disc is dispersed is generally well beyond feasibility for 2D hydrodynamical calculations. What is done instead is to use the 2D steady-state solution to determine a one-dimensional mass-loss profile, $\dot{\Sigma}_\mathrm{w}(R)$ \citep{Clarke2001,Matsuyama2003,AlexanderArmitage2009,Monsch2021}, which is then included as a sink term to the one-dimensional disc viscous evolution equation \citep{LyndenBellPringle1974,Pringle1981} to study the surface density evolution of the disc as a function of time:

\begin{equation} 
\frac{\partial \Sigma}{\partial t} = \frac{1}{R}\frac{\partial}{\partial R}\left[ 3R^{1/2} \frac{\partial}{\partial R}\left(\nu \Sigma R^{1/2}\right) \right] - \dot{\Sigma}_{\mathrm {w}}(R,t).
\label{eq:sigma_evol}
\end{equation}

\begin{figure}
    \centering
    \includegraphics[width=\textwidth]{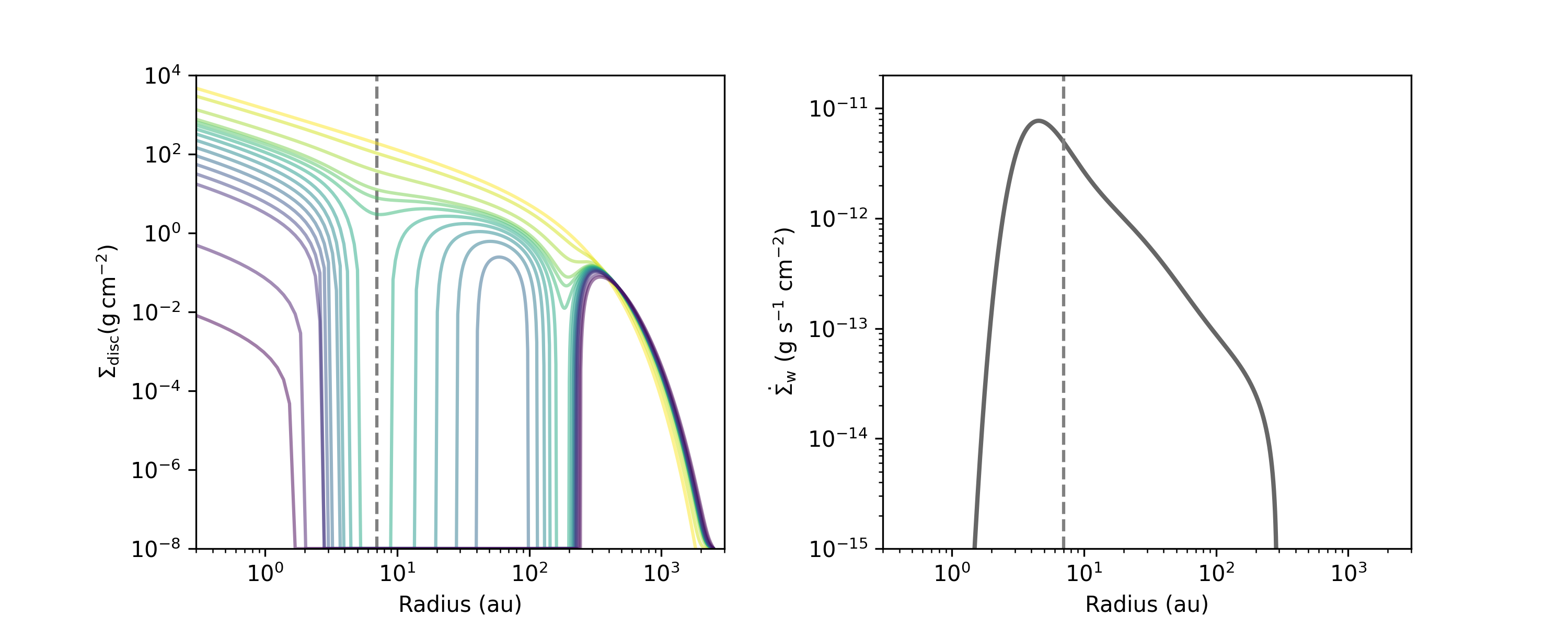}
    \caption{On the left panel, the 1D surface density evolution as a function of disc radius is shown for a disc of $M_d = 0.1 M_\odot$, orbiting a Solar mass star, with $L_X = 2.04\cdot10^{30}$ erg s$^{-1}$ \citep{Picogna2021}. The different lines are drawn at [0, 25, 50, 60, 62, 64, 66, 68, 70, 72, 74, 76, 78, 80, 90, 99] \% of the corresponding total disc lifetime. The dotted line shows the approximate location of the inner gap opening due to photoevaporation. On the right panel, the correspoding Surface density mass-loss rate due to photoevaporation is shown as a function of disc radii.}
    \label{fig:sigma_evol}
\end{figure}

Figure~\ref{fig:sigma_evol}, shows the mass-loss profile for a 1 M$_{\odot}$ (right panel \citep{Picogna2021}) and the corresponding surface density evolution (left panel). The surface density evolution (time and location of gap opening, dispersal timescale etc.) is extremely sensitive to $\dot{\Sigma}(R)$, which in turn affects also the final architecture of planets formed in the disc \citep{Alexander2012,ErcolanoRosotti2015,Jennings2018}, as well as the potential of photoevaporation of triggering the formation of planetesimals by the streaming instability \citep{Carrera2017,Ercolano2017}. 

\subsection{Divergence of the results}
\label{sec:divergence}

While the broad-brush picture described above is shared by most authors, the $\dot{\Sigma}_\mathrm{w}(R)$, and the physical properties of the wind derived by different projects (e.g. see Table 1) diverge significantly.
A recent comparison of the Surface density mass-loss between the different models of disc photoevaporation \citep[][see their Figure~7]{Pascucci2022} has highlighted an order of magnitude difference in the cumulative mass-loss rates which depends not only on the location of the surface mass-loss rate maximum but also on the extent of the profiles. These differences can be broadly understood looking at the choices made for the frequency range adopted for the stellar irradiation, and the different methods used. However, most of them present a peak inside 10 au and a slow decline of the surface density mass-loss rate in the outer regions.

As photoevaporation is driven by stellar irradiation, a large part of the divergence can be understood by considering the input spectrum assumed by different authors (see also discussion in \citep{Lesur2022}, section 4.3.1). A pure EUV model \citep[e.g.][]{Alexander2006a,Alexander2006b} yields an almost isothermal gas with a temperature around $10^4\,\mathrm{K}$. In this case the mass-loss peaks around the gravitational radius which is about $9$ au for a 1 $M_{\odot}$ star, the total mass-loss scales as the square root of the EUV flux, and it is roughly 10$^{-10}$ M$_{\odot}$/yr assuming a EUV flux of 10$^{41}$ phot/sec \citep{Alexander2006a, Alexander2006b}. 

Soft X-ray radiation penetrates deeper in the disc than EUV radiation and yields mass-loss rates that are one or two orders of magnitude higher than the classical EUV model, depending almost linearly on the X-ray luminosity of the irradiating star \citep{Owen2012,Picogna2019,Picogna2021}. X-ray photoevaporation models, all include a EUV component, but find it is irrelevant to driving the wind \citep[e.g.][]{Ercolano2009}. Models for 1 M$_{\odot}$ star yield total mass-loss rates of order 10$^{-8}M_{\odot}/yr$ assuming a soft spectrum and X-ray luminosities of 10$^{30}$ erg/sec \citep{Owen2010, Picogna2019}. The X-ray-driven mass-loss profile is much more extended than in the EUV case, peaking around the gravitational radius, but extending out to $\sim 200$ au. 
Carbon is one of the major contributors in the X-ray opacity, since X-ray photons are mainly absorbed by the inner shells of the more abundant heavy elements in the gas and dust \citep{ErcolanoClarke2010}. Carbon depletion is expected as a natural consequence of disk evolution both chemical as the Carbon turns from CO into more complex species, and physical due to grain growth that locks up large fractions of Carbon in ice bodies. For Carbon depleted disks, the magnitude and extent of the mass-loss rates are expected to increase by a factor $\sim 2$ \citep{Woelfer2019}.
Internal FUV radiation is expected to drive photoevaporation from further out in the disc, and initial estimates, based on the hydrostatic equilibrium models of \cite{Gorti2009}, found it to be very efficient under the assumption that Polycyclic Aromatic Hydrocarbons (PAHs) are abundant in the atmosphere of discs. However, the atmospheric abundance of PAHs, which are rarely observed in T-Tauri discs \citep{Seok2017}, is likely to be very small, furthermore stellar FUV flux depends on the accretion rate onto the central star and decreases with time. For these reasons, the role of FUV-driven winds on the final disc dispersal is uncertain.

More recent calculations \cite{Nakatani2018a}, while considering EUV, FUV and X-ray simultaneously, come to the conclusion that \emph{direct} EUV flux dominates the driving of the wind, but obtain two orders of magnitude higher mass-loss rates than previous EUV-only models. These calculations are in direct contrast with the classical picture that stellar (direct) EUV photons are absorbed in the bound inner regions of the disc and the diffuse field coming from the puffed-up inner disc drives photoevaporation (at low rates) at the gravitational radius. The assumption that direct EUV are able to reach the disc at and beyond the gravitational radius relies on previous calculations \citep{Tanaka2013}. The source of the discrepancy between these more recent calculations and previous works  might lie in the specific disc geometries assumed, where the inner disc does not expand enough to screen the outer regions or where there is significant flaring, which might not apply to the majority of discs \citep{Hollenbach2017}. We finally note that these results are also in tension with previous work \citep{Ercolano2009} which by means of detailed 2D Monte Carlo RT and thermal calculation in a hydrostatic disc, explicitly test the role of EUV in the presence of an observationally derived X-ray spectrum and find EUV effects to be negligible on the final mass-loss rates. A detailed comparison in a dedicated work must however be carried out to finally determine the nature of this discrepancy.                                            %\textbf{They adopted an observationally derived X-ray spectrum, but targeted on a specific system (TW Hydrae) which has a larger spectral hardness with respect to the average T Tauri population, causing lower efficiency of X-ray photoevaporation \citep{Ercolano2009}.}

Strong divergence is also observed in the physical properties of the wind obtained by recent calculations  \citep{WangGoodman2017} which use the \textsc{Athena++} code \citep{White2016,Stone2020} to perform self-consistent radiation hydrodynamics and thermochemical calculation of photoionised disc compared to models that use the $\xi-T_e$ approach  \cite{Owen2010,Owen2012,Picogna2019,Picogna2021, Ercolano2021}. While total mass-loss rates obtained are comparable, the \textsc{Athena++} models come to the conclusion that X-ray radiation is inefficient at driving the wind and that EUV dominates. Also the temperature structure of the wind is significantly different in the two cases. The \textsc{Athena++} models predict extremely high temperatures ($10^5$ K) at the base of the wind which decreases higher up, in contrast with the Parker-like wind obtained by the $\xi-T_e$ models, which has a few $10^3$ K launching temperature and reaches top temperatures of $10^4$ K but only for the very limited region of the wind that is heated (not launched!) by EUV radiation. The divergence of these models has been often attributed to methodological differences on how to approach to radiative transfer and thermal balance, the choice of irradiating spectrum employed, and the processes available to cool the gas.
A recent detailed comparison \citep{Sellek2022} has been able to shed light on this question,  demonstrating that the divergence between these models is driven predominantly by the choice of the irradiating spectrum and the very limited number of frequency points used to describe the radiation field by the \textsc{Athena++} model (7) compared to the $\xi-T_e$ models ($>$1000).  The $\xi-T_e$ models use an X-ray spectrum derived from deep Chandra observations of T-Tauri stars \citep{Getman2005, Ercolano2008b, Ercolano2021}, while the \textsc{Athena++} model assumes an analytical distribution.

\section{Observational tests}

For complex theoretical models to become a realistic description of natural phenomena rather than remaining numerical experiments, it is very important that they provide predictions that can be tested against observations. 
In this Section, we list a number of direct and indirect tests that have been used to constrain theoretical photoevaporation models. 

\subsection{Direct Tests}
\begin{figure}
    \centering
    \includegraphics{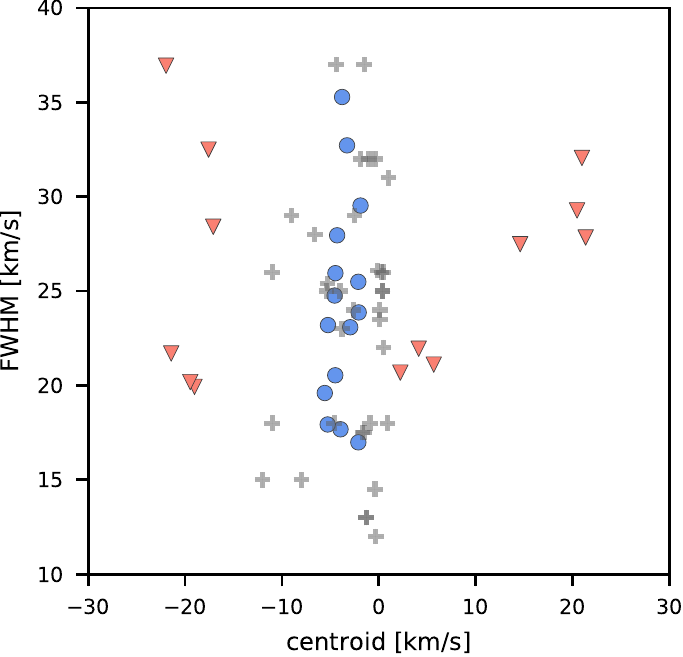}
    \includegraphics{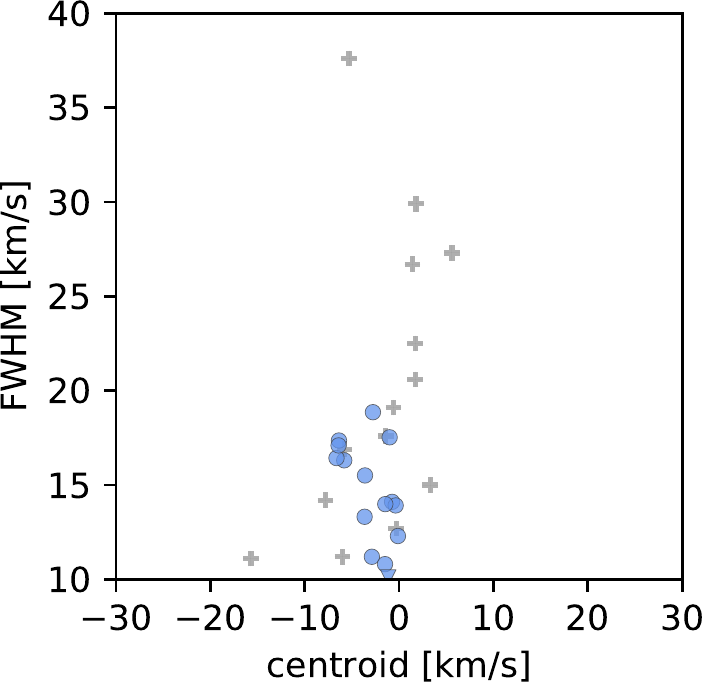}
    \caption{Left Panel: Full width at half maximum vs peak velocity of the narrow low-velocity components (NLVC) which result from the multi-Gaussian decomposition of [OI] 6300 \AA{} line profiles. Grey plusses mark the observed NLVCs \citep{Banzatti2019}. Blue circles and red triangles show NVLCs from the photoevaporation and MHD wind models \citep{Weber2020}, respectively. Right Panel: Same as Fig. 2 but for the NLVC of the $\mathrm{o\!-\!H_2}\,1\!-\!0\,\mathrm{S}(1)$ at $2.12\;\mathrm{\mu m}$. Grey pluses show the observed NLVCs [70]. The blue circles show the results for the same physical photoevaporative wind models as shown in Fig. 2, but post-processed with a thermo-chemical code to properly model the molecular hydrogen chemistry and excitation [Rab et al., subm.]. }
    \label{fig:fwhm}
\end{figure}

Winds are tenuous and direct imaging of its gas component is not possible. However, the gaseous component of the wind can still be "observed" directly via high-resolution spectroscopy of emission lines that are emitted in the outflow and show a blue-shifted component \citep{ErcolanoPascucci2017,Pascucci2022}. Line profiles of collisionally ionised lines of neutral and low ionisation species ([OI], [NeII], [SII], [N2], [FeII]), as well as some molecular tracers (CO and H$_2$), are available for a statistically significant number of T-Tauri stars \citep[e.g.][]{Simon2016,Banzatti2019,Gangi2020}. The majority of the observed profiles can be well-fitted by X-ray-driven photoevaporative models \citep{ErcolanoOwen2010,ErcolanoOwen2016,Rab2022}, in combination with a magnetically driven component for sources showing composite profiles \citep{Weber2020} (see Figure~\ref{fig:fwhm}).
Lines in the sub-mm region (e.g. CO and CI) have also been used to infer the presence of disc winds in highly inclined sources \citep{Louvet2018} and some predictions from numerical MHD wind models already exist \citep{Gressel2020}.
%{\color{red} shall we include a figure here showing a fit of H2 from christian's paper and a composite profile from michael's paper?}

Ionised gas in the wind and disc atmosphere region could also be detected and spatially constrained via its free-free emission at cm-wavelengths using observations from upcoming high spatial resolution facilities like ngVLA \citep[e.g.][]{Ricci2021}. This is a promising avenue to distinguish magnetic winds which can be driven from regions much closer to the star than photoevaporative winds. 

Small ($\lessapprox$ 10$\mu$m) dust grains can be entrained in the wind from the launching region \citep{Owen2011,Ercolano2011,Hutchison2016,Hutchison2017,Franz2020,Hutchison2021,Booth2021,Rodenkirch2022}, and it may be possible to detect their signature in scattered light observations of highly inclined sources, particularly from discs with inner dust cavities \citep{Franz2022a,Franz2022b}.

\subsection{Indirect Tests}

Indirect constraints that photoevaporation models should match might include (i) observed disc dispersal timescales which can be traced by the evolution of the accretion properties and surface density of observed disc populations, including during the transition disc phase \citep[e.g.][]{Luhman2010,Ercolano2011,Ercolano2015,Koepferl2013,Rosotti2013,Ercolano2018,Garate2021}; (ii) metallicity dependence of disc lifetimes \citep{Yasui2009,Yasui2010,Yasui2014,ErcolanoClarke2010,Takagi2015,Nakatani2018a}; (iii) observed correlations in disc populations like the $\dot{M}-M_\star$ relation \citep[e.g.][]{Ercolano2014,Sellek2020, Somigliana2020}.

It is important to note that in order to test photoevaporation models against observed disc populations, wind properties (mass-loss rates as a function of disc radius) must be calculated for a wide enough parameter space, spanning the observed stellar masses and X-ray properties. Assumptions are then necessary as to how angular momentum is transported in the disc. Most studies to date have assumed a viscous model in combination with photoevaporation. It would be interesting to repeat this body of work assuming angular momentum transport by magnetised disc winds \citep[e.g.][]{Tabone2022,Trapman2022}.

\section{Outlook}

As mentioned at the end of the last section, magnetised disc winds might be present in combination with thermal winds and the former might provide the dominant mechanism for angular momentum transport instead of viscosity. Direct hints of their presence are seen in emission line profiles observed for some sources which show multiple outflow components \citep[e.g.][for a recent review]{Pascucci2022}. Furthermore non-ideal magnetohydrodynamical models routinely observe magnetised winds while they struggle to maintain magnetorotational instability active in most regions of the disc \citep[e.g.][for a recent review]{Lesur2022}. Clearly the frontier of modelling disc winds is to account for both processes simultaneously and at high enough spatial resolution to produce models that have enough predictive power to be confronted with the observations. 
Initial attempts are promising \citep[e.g.][]{Wang2019,Rodenkirch2020}, but they still lack resolution in the inner disc regions (where most observed diagnostics come from), have a large lower density threshold (comparable to the wind density, making post-processing of the winds problematic), oversimplify radiative transfer and rely on a number of unconstrained parameters to describe the magnetic flux and its evolution. If these problems can be overcome these models could acquire enough predictive power for a meaningful comparison with the observations, as detailed in the previous section, and thus provide us with  powerful tools to understand how protoplanetary discs evolve and finally disperse. 

\section{Acknowledgements}
We thank the anonymous referee for a thorough and constructive report that helped improve this review.
We acknoweledge the support of hte Excellence Cluster ORIGINS which is funded by the Deutsche Forschungsgemeinschaft
(DFG, German Research Foundation) under Germany´s
Excellence Strategy – EXC-2094-390783311. BE, GP and CHR
acknowledge the support of the Deutsche Forschungsgemeinschaft
(DFG, German Research Foundation) - 325594231

\begin{sidewaystable}
\footnotesize
\sidewaystablefn%
\begin{center}
\begin{minipage}{\textheight}
\caption{\textsc{Model comparison}}\label{tab3}
\begin{tabular*}{\textwidth}{@{\extracolsep{\fill}}c|ccccc@{\extracolsep{\fill}}}
\toprule
\textsc{Project}            & \textsc{Hydro}    & \textsc{Dust}                 & \textsc{Radiation} & \textsc{Chemistry} & \textsc{Spectrum} \\
\toprule
\textsc{Wang et al.}$^{1}$      & yes               & $r_\mathrm{dust}:5 \AA$       & yes                           & 24 species of H,He,C,O,S,     & 4 energy bins: 7 eV (FUV),\\
                                &                   & d/g: $7\cdot10^{-5}$          & \textit{ray-tracing}          & Si,Fe,dust grains,e$^{-}$,    & 12 eV (L–W), 25 eV (EUV), \\
                                &                   & \textit{no dynamics/growth}   & \textit{(no diffuse field)}   & $\sim 100$ chemical reactions & and 1 keV (X-ray band)    \\
\midrule
\textsc{Nakatani et al.}$^{2}$  & yes   & MRN ($3.1 \AA \to 0.01 \mu m$)     & yes   & 8 species of H,C,O, grains    & $81$ freq. bins for FUV/EUV\\
                                &       & d/g: $10^{-6}\to10^{-1}$  & \textit{ray-tracing} & 13 chemical reactions         & X-ray SED derived from\\
                                &       & \textit{no dynamics/growth} & \textit{(no diffuse field)}      &                               & TW-Hya\\
\midrule
$\xi-T_e$                       & yes   & MRN ($50 \AA \to 0.25\mu m$) & yes                               & no    & $\gt 1000$ freq. bins\\
\textsc{approach}$^3$           &       & d/g: $2.5,4\cdot10^{-4}$  & \textit{Monte Carlo RT}    &       & \textit{obs. derived input spectra}\\
                                &       & \textit{no dynamics/growth} & \textit{(radiative equilibrium)} & & \textit{different spec. hardness}\\
\midrule
\textsc{Gorti et al.}$^{4}$   & no                            & MRN ($50 \AA \to 20\mu m$) & yes                           & 84 species of H, He, C, O,    &\\
        & \textit{vertical hydrostatic} & d/g: $0.01$  & \textit{1+1D gas and dust}    & Ne,S, Mg, Fe, Si, Ar, S, Mg,  & L$_X$(E)$\propto$ E [$0.1 - 2$] keV\\
        & \textit{equilibrium}          & \textit{no dynamics/growth}  & \textit{radiative transfer}   & Fe, Si, Ar,                   & $L_X(E)\propto E^{-1.75}$ [$2 - 10$] keV \\
        &                               &   &                               &  $\sim 600$ chemical reactions & \\
\midrule
\textsc{Ercolano et al.}$^{5}$   & no                            & MRN ($50 \AA \to 0.25\mu m$)  & yes                                   & atomic and photoionised               & $\gt 1000$ freq. bins, \\
        & \textit{vertical hydrostatic} & d/g: $2.5,4\cdot10^{-4}$  & \textit{full Monte Carlo RT}          & species of H, He, C, O, Ne,           & \textit{obs. derived input spectrum}\\
        & \textit{equilibrium}                       & \textit{no dynamics/growth}  & with dust and gas                     & S, Mg, Fe, Si, Ar & \\
\midrule
\textsc{Alexander et al.}$^{6}$   & yes &  & no & no & -- \\ 
\botrule
\end{tabular*}
\footnotetext{$^{1}$\citep{WangGoodman2017,Wang2019},$^2$\citep{Nakatani2018a,Nakatani2018b,Komaki2021},$^3$\citep{Owen2010,Owen2012,Picogna2019,Woelfer2019,Ercolano2021,Picogna2021},$^4$\citep{Gorti2009},$^5$\citep{Ercolano2009},$^6$\citep{Alexander2006a,Alexander2006b}}
\end{minipage}
\end{center}
\end{sidewaystable}

Data Availability Statement: No Data associated in the manuscript

\bibliography{sn-bibliography}% common bib file


\end{document}